\documentclass[
    ,final            
  ]
  {aipproc}

\layoutstyle{8x11double}

\usepackage{amssymb}

\begin{document}

\title{Polar kicks and the spin period -- eccentricity relation in double neutron stars}


\author{B. Willems}{
  address={Northwestern University, Department of Physics and Astronomy,
  2131 Tech Drive, Evanston, IL 60208, USA}
}

\author{J. Andrews}{
  address={Northwestern University, Department of Physics and Astronomy,
  2131 Tech Drive, Evanston, IL 60208, USA}
}

\author{V. Kalogera}{
  address={Northwestern University, Department of Physics and Astronomy,
  2131 Tech Drive, Evanston, IL 60208, USA}
}

\author{K. Belczynski}{
  address={New Mexico State University, Department of Astronomy, 1320 Frenger Mall, Las Cruces, NM 88003, USA}
  ,altaddress={Tombaugh Fellow} 
}

\begin{abstract}
We present results of a population synthesis study aimed at examining the role of spin-kick alignment in producing a correlation between the spin period of the first-born neutron star and the orbital eccentricity of observed double neutron star binaries in the Galactic disk. We find spin-kick alignment to be compatible with the observed correlation, but not to alleviate the requirements for low kick velocities suggested in previous population synthesis studies. Our results furthermore suggest low- and high-eccentricity systems may form through two distinct formation channels distinguished by the presence or absence of a stable mass transfer phase before the formation of the second neutron star.  The presence of highly eccentric systems in the observed sample of double neutron stars may furthermore support the notion that neutron stars accrete matter when moving through the envelope of a giant companion.
\end{abstract}


\maketitle


\section{Introduction}
\begin{figure}
  \includegraphics[height=.24\textheight]{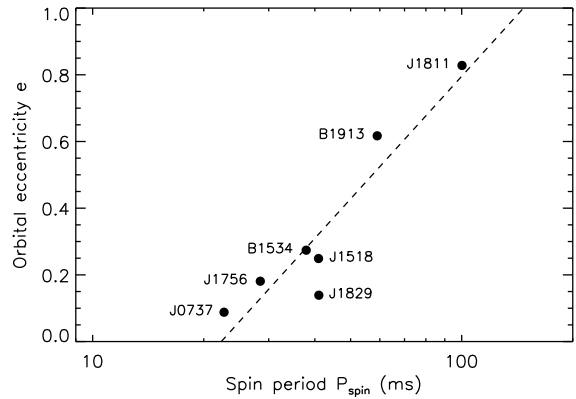}
  \caption{Orbital eccentricities and spin periods of observed DNSs in the Galactic disk. The dashed line represents the best-fitting log-linear curve $e=-1.7 + 1.2 \log P_{\rm spin}$.}
  \label{f1}
\end{figure}

Recent observations of single and binary pulsars have sparked new questions and challenged accepted ideas on the formation of neutron stars (NSs) and the nature of supernova (SN) kicks. In particular, measurements of pulsar radio emission polarization and pulsar wind nebulae symmetry axis directions have provided increasingly compelling evidence for the alignment of pulsar proper motions and pulsar rotation axes \citep[e.g.][]{2005MNRAS.364.1397J, 2006ApJ...639.1007W, 2006ApJ...644..445N, 2007ApJ...660.1357N, 2007ApJ...664..443R, 2007arXiv0708.4251J}.
Assuming the proper motion and pulsar spin axis directions are representative of the natal kick and NS progenitor rotation axis, the alignment suggests that natal kicks are preferentially aligned with the progenitor's rotation axis. Moreover, the increasing sample of double neutron star (DNS) binaries has revealed a possible correlation between the spin period $P_{\rm spin}$ of the first born NS and the binary orbital eccentricity $e$ (see Fig.~\ref{f1}) \citep{2005ASPC..328...43M, 2005ApJ...618L.119F}. Such a relation arises naturally for symmetric SN explosions, but is highly constraining for asymmetric explosions. Our aim in this paper is to examine the role of spin-kick alignment in establishing the observed $P_{\rm spin}$--$e$ correlation.

\section{Double neutron star formation}

According to our current understanding, DNSs form from binaries in which, possibly after some mass exchange, both stars are massive enough to undergo core-collapse into a NS. After the formation of the first NS, the binary evolves through a high-mass X-ray binary phase during which the NS accretes matter from the wind of its companion. When the companion exhausts its nuclear fuel, it swells up to giant dimensions and initiates a dynamically unstable mass transfer phase. As the NS moves through the expanding envelope, orbital energy and angular momentum are transferred to the envelope, spinning it up until it is expelled from the system (assuming enough orbital energy is available for the expulsion). The binary emerges from this "common envelope" phase consisting of a NS and the helium core of its former giant companion. If the NS is able to accrete matter while moving through the common envelope, the associated accretion of angular momentum may spin it up to subsecond rotation periods. After the common-envelope phase, the giant's exposed helium core evolves further until it, in turn, explodes in a SN and forms a second NS. Depending on the mass of the helium core and the size of the orbit right before the SN explosion, the formation of the second NS may be preceded by a dynamically stable mass transfer phase during which accretion spins-up the first-born NS to millisecond periods. This formation channel is commonly referred to as the standard DNS formation channel \citep{1991PhR...203....1B}.

The orbital characteristics of observed DNS binaries are determined by the mass loss and SN kick accompanying the birth of the second NS, and the subsequent loss of orbital energy and angular momentum through gravitational radiation. In the case of an asymmetric SN explosion, the orbital semi-major axis $A$ and eccentricity $e$ right after the explosion depend on the mass loss and kick imparted to the NS through the conservation laws of orbital energy and angular momentum \citep[e.g.][]{1983ApJ...267..322H, 1995MNRAS.274..461B, 1996ApJ...471..352K, 1997ApJ...489..244F, 2000ApJ...530..890K}: 
\begin{eqnarray}
\lefteqn{V_{\rm k}^2 + V_0^2 
 + 2\, V_{\rm k}\, V_0\, \cos
 \theta} \nonumber \\
 & & = G \left( M_1 + M_2 \right) \left( {2 \over A_0}
 - {1 \over A} \right),  \label{eq1}
\end{eqnarray}
\begin{eqnarray}
\lefteqn{A_0^2 \left[ V_{\rm k}^2\, \sin^2 \theta \cos^2
 \phi \right. + \left. \left( V_{\rm k}\, \cos \theta 
 + V_0 \right)^2 \right]} \nonumber \\
 & = & G \left( M_1 + M_2 \right) A 
 \left( 1 - e^2 \right). \hspace{1.2cm} \label{eq2}
\end{eqnarray}
Here $G$ is the Newtonian gravitational constant, $M_1$ and $M_2$ are the masses of the first- and second-born NS, $V_0=[G(M_1+M_0)/A_0]^{1/2}$ is the relative orbital velocity of the second NS's progenitor right before its SN explosion, $M_0$ is the pre-SN mass of the second NS's progenitor, $A_0$ is the pre-SN orbital separation, and $V_k$ is the magnitude of the natal kick velocity. The angles $\theta$ and $\phi$ define the direction of the natal kick velocity: $\theta$ is the polar angle between $\vec{V}_k$ and $\vec{V}_0$, and $\phi$ the corresponding azimuthal angle in the plane perpendicular to $\vec{V}_0$ (with $\phi=\pi/2$ corresponding to the direction from the first NS to the progenitor of the second NS). We note that in writing down Eqs.~(\ref{eq1})--(\ref{eq2}), the pre-SN orbit was assumed to be circular, as expected from the strong tidal forces acting during the common envelope occurring in the course of DNS formation.

In the case of a symmetric SN explosion, no natal kick is imparted to the NS and the post-SN orbital elements are uniquely determined by the mass loss from the system. Equations~(\ref{eq1})--(\ref{eq2}) reduce to \citep[e.g.][]{1961BAN....15..265B,1961BAN....15..291B} 
\begin{equation}
{A \over A_0} = {{M_1+M_2} \over {M_1+2\,M_2-M_0}}, \label{sym1}
\end{equation} 
\begin{equation}
e = {{M_0 - M_2} \over {M_1 + M_2}}. \label{sym2}
\end{equation}

\section{The $P_{\rm spin}-e$ relation}

\begin{figure}
  \includegraphics[height=.24\textheight]{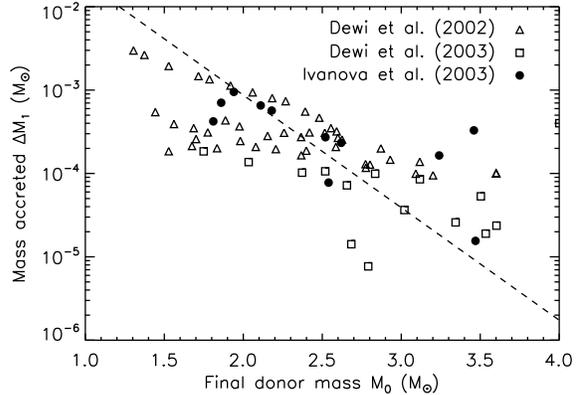}
  \caption{Mass accreted by the NS as a function of the helium star mass at the end of a stable mass transfer phase in NS--helium star binaries (data from \citep{2002MNRAS.331.1027D, 2003MNRAS.344..629D, 2003ApJ...592..475I}). The amount of mass accreted increases with decreasing donor star mass because the duration of the mass transfer phase increases with decreasing donor mass. The dashed line represents the best-fitting log-linear curve $\log \Delta M_1 = -0.36 - 1.35\, M_0$.}
  \label{f2}
\end{figure}

\citet{2005ApJ...618L.119F} noted that, in the absence of SN kicks, a relation between $P_{\rm spin}$ and $e$ arises naturally if the degree of spin-up of the first-born NS correlates with the mass of the progenitor of the second-born NS. Such a correlation is expected for binaries in which a stable mass transfer phase precedes the SN explosion of the progenitor of the second NS. The reason for this is twofold. First, less massive stars evolve more slowly than more massive stars, yielding longer-lived mass-transfer phases and thus more time to accrete matter and spin the first NS up to shorter spin periods (see Fig.~\ref{f2}). Second, in the absence of SN kicks, the binary orbital eccentricity after the SN explosion forming the second NS is proportional to the amount of mass lost during the SN (see Eq.~\ref{sym2}). Less massive progenitor stars therefore not only imply more spin-up but also lower orbital eccentricities. \citet{2005MNRAS.363L..71D} have shown this theoretically expected correlation to persist if small kicks of the order of a few 10\,km\,s$^{-1}$ are imparted to the second NS at birth. Larger kicks of a few $100\,{\rm km\,s^{-1}}$, typical for isolated radio pulsars, spread out the range of post-SN orbital eccentricities, destroying any correlation between the first NS's spin period and the binary orbital eccentricity. 

The requirement of small kicks to reproduce the observed $P_{\rm spin}$--$e$ correlation is puzzling considering the stringent lower limits on the kick velocities required to explain the observed binary properties and space velocities of the relativistic DNS binaries PSR\,B1534+12 and PSR\,B1913+16  (see \citep{2004ApJ...616..414W, 2005ApJ...619.1036T} and the paper by I. Stairs in this volume). Yet, these binaries follow the $P_{\rm spin}$--$e$ correlation shown in Fig.~\ref{f1}. A revision of the low kick scenario proposed to explain the $P_{\rm spin}$--$e$ relation is therefore in order.

\section{Polar kicks}

In the absence of any theoretical models arguing otherwise, the formation of single and binary NSs has so far primarily been studied assuming all kick directions to be equally likely (some tentative considerations of non-isotropic kicks are presented in \citep{2004ApJ...616..414W, 2000ApJ...541..319K}). However, in recent years, observational evidence that NS kicks may be preferentially aligned with NS rotation axes has become increasingly convincing: evidence for alignment between pulsar proper motion directions and pulsar wind nebula symmetry axes was presented by \citet{2006ApJ...639.1007W} and \citet{2006ApJ...644..445N, 2007ApJ...660.1357N}, while \citet{2005MNRAS.364.1397J, 2007arXiv0708.4251J} and \citet{2007ApJ...664..443R} presented evidence for alignment between pulsar proper motion directions and polarization vectors. Assuming the proper motion directions are representative of the kick directions, these alignments suggest a correlation between pulsar rotation axes and kick directions. In what follows, we furthermore assume the rotation axes of the pulsar and its progenitor to be parallel, so that the above evidence can also be considered as evidence for alignment between NS kick directions and NS progenitor rotation axes. We will refer to such kicks as "polar kicks".

If SN kicks are restricted to be along the rotation axis of the NS progenitor ($\theta=\pi/2$ and $\phi=0$ or $\pi$), Eqs.~(\ref{eq1})--(\ref{eq2}) reduce to
\begin{equation}
A = \frac{A_0}{1 \pm e},  \label{eq3}
\end{equation}
\begin{equation}
e = \frac{M_0-M_2}{M_1+M_2} \left[ 1 + \frac{A_0}{G(M_0-M_2)} V_k^2 \right].  \label{eq4}
\end{equation}
Hence, the post-SN orbital eccentricity is approximately proportional to the amount of mass lost during the SN explosion if the spread in the values of the second term between the square brackets in Eq.~(\ref{eq4}) is sufficiently small. The question therefore arises whether typical values of $A_0/[G(M_0-M_2)]$ for DNS progenitors are small enough to relax the requirement for low kicks to explain the observed $P_{\rm spin}$--$e$ relation.

\section{Population synthesis}

To examine the role of polar kicks in establishing the $P_{\rm spin}$--$e$ relation, we use the StarTrack binary population synthesis code to construct populations of DNSs forming through the standard evolutionary channel (see \citep{2002ApJ...572..407B, 2005astro.ph.11811B} for an extensive and detailed description of the population synthesis code).  In this analysis, we consider population synthesis models with the same input parameters as the standard model (model A) of \citet{2005astro.ph.11811B}, except that we do not allow for hypercritical accretion onto the NS during common envelope evolution and we modify the adopted kick velocity distribution to explore the effects of polar kicks. In particular, the first-born NS is assumed to be born with a kick velocity typical of isolated radio pulsars drawn from a Maxwellian velocity distribution with dispersion $\sigma=300\,{\rm km\,s^{-1}}$ \citep{2005MNRAS.360..974H, 2005MNRAS.362.1189Z}, while the second-born NS is assumed to be born with a kick velocity drawn from a Maxwellian velocity distribution with dispersion $\sigma=20\,{\rm km\,s^{-1}}$ (model~K20) or $\sigma=300\,{\rm km\,s^{-1}}$ (model~K300). The direction of the kick imparted to the first-born NS is furthermore assumed to be isotropic, while the direction of the kick imparted to the second-born NS is constrained to be within $10^\circ$ from the progenitor's rotation axis.

In the context of the standard evolutionary channel, the first-born NS is spun-up by dynamically stable mass transfer from the stripped down helium core of the NS's companion. The accretion rate is at all times limited to the Eddington rate for helium-rich matter. Spin-up is then modeled assuming the NS has an initial spin period of 1\,s at the start of the mass transfer phase and accretion of a mass $\Delta M_1$ increases NS's spin angular momentum by
\begin{equation}
\Delta J_{\rm spin} = \sqrt{G\,M_1\,R_{\rm A,1}}\, \Delta M_1,
\end{equation}
where $R_{\rm A,1}$ is the Alfv\'en radius of the first-born NS (see also \citep{2005MNRAS.363L..71D}). We calculate the latter by randomly drawing a magnetic field strength from a uniform distribution between $10^9$ and $10^{10}$\,G. 

\begin{figure*}
  \includegraphics[height=.24\textheight]{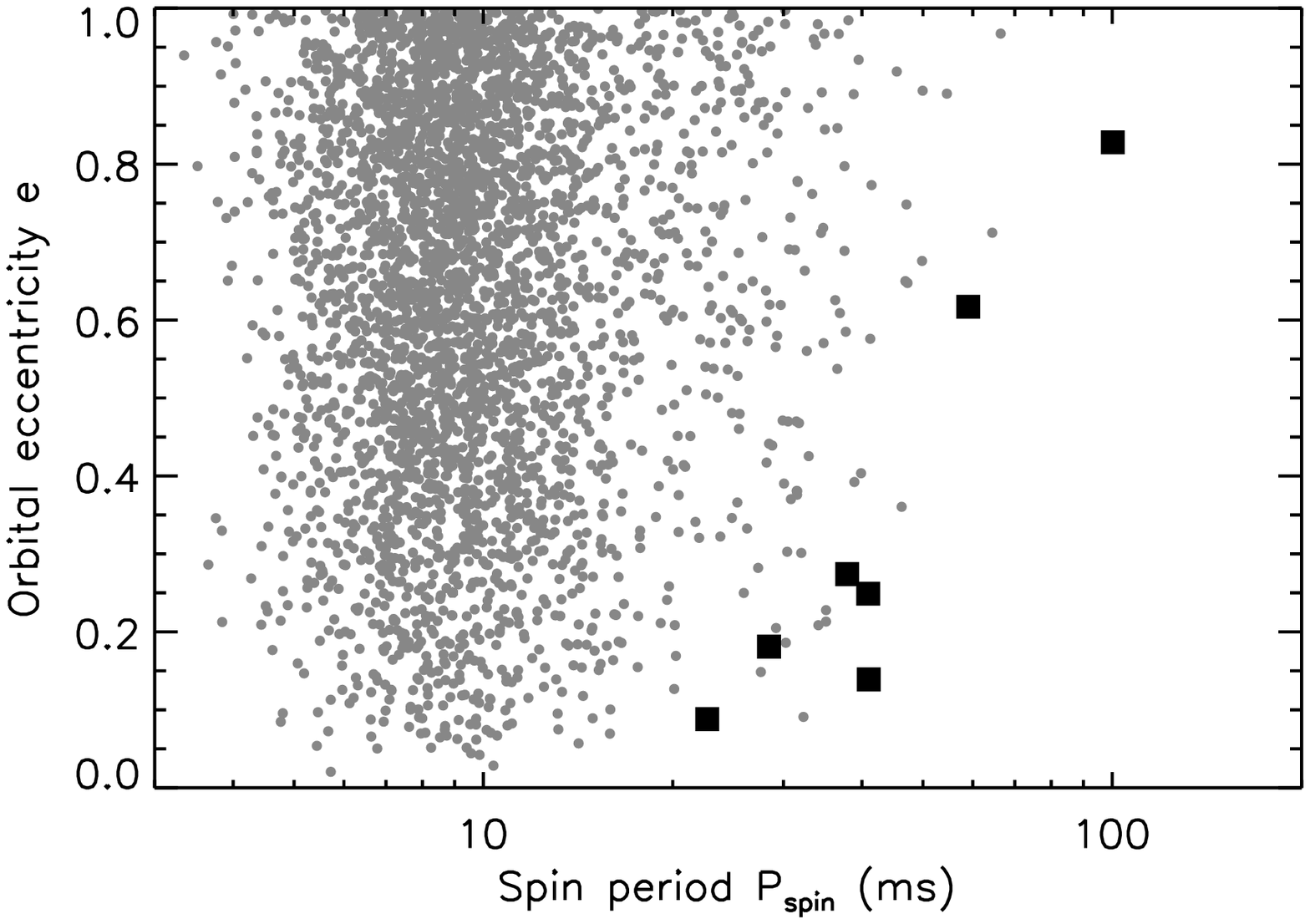}
  \includegraphics[height=.24\textheight]{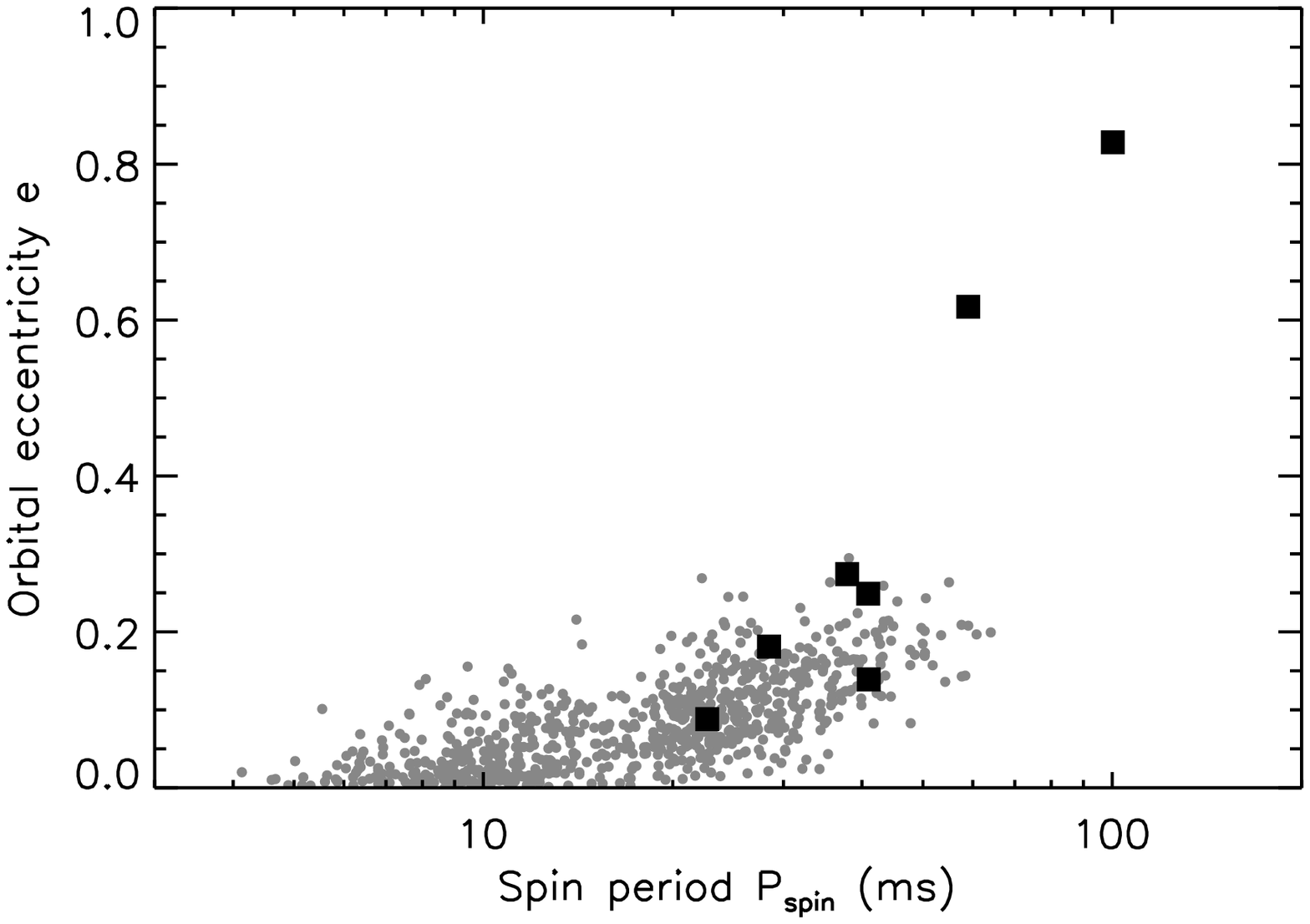}
  \caption{Scatterplots showing the present-day orbital eccentricities and spin-periods of the first-born NS in DNS binaries for the considered SN kick models (left: model K300; right: model K20). The present-day orbital eccentricities are determined by evolving the post-SN orbital elements under the influence gravitational wave emission up to the present epoch. Squares indicate the orbital eccentricities and spin-periods of the first-born NSs of observed Galactic DNSs.}
  \label{f3}
\end{figure*}

A scatterplot of the present-day orbital eccentricities and spin-periods of the first-born NS in the resulting DNS populations is shown in Fig.~\ref{f3}. In the model where kicks typical of isolated radio pulsars (model~K300) are imparted to the second-born NS, no correlation appears between $P_{\rm spin}$ and $e$. However, when low kicks are imparted to the second-born NS (model~K20), the low-$e$ end of the correlation is well sampled by the simulations, although a significant number of systems is also formed with spin periods $P_{\rm spin} < 20$\,s below those of the observed systems. A possible cause may be the oversimplified nature of our NS spin-up model. These conclusions are in agreement with those of \citet{2005MNRAS.363L..71D} who studied the $P_{\rm spin}$--$e$ correlation assuming isotropic kick distributions. 

\begin{figure}
  \includegraphics[height=.24\textheight]{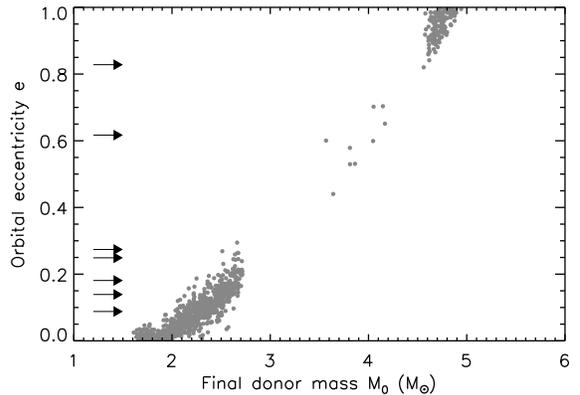}
  \caption{Scatterplot showing the present-day orbital eccentricities and progenitor masses of the second-born NS in DNS binaries for SN kick model K20. The arrows on the left indicate the orbital eccentricities of observed Galactic DNSs.}
  \label{f4}
\end{figure}

As can be seen from Fig.~\ref{f3}, the main problem of the low kick scenario is its inability to explain the two observed DNS binaries PSR\,B1913+16 and PSR\,J1811-1736 with eccentricities $e \gtrsim 0.6$. The problem arises from the low NS progenitor masses ($M_0 \lesssim 2.7\,M_\odot$) at the end of the mass transfer phase spinning up the first NS: for $M_0 \lesssim 2.7\,M_\odot$ and $M_1=M_2=1.35\,M_\odot$, the maximum attainable orbital eccentricity without a SN kick is 0.5. In the absence of kicks, high-$e$ systems are therefore naturally filtered out by the standard evolutionary channel\footnote{\citet{2005MNRAS.363L..71D} found a significant number of high-eccentricity systems in their population synthesis study of DNS binaries based on isotropic SN kicks with low kick velocities. While it is tempting to ascribe this difference to the different formation channels considered in our and their paper, both channels rely on spin-up of the first-born NS during a dynamically stable mass transfer phase prior to the SN explosion forming the second NS. The origin of the high-eccentricity systems in the absence of significant SN kicks in the \citet{2005MNRAS.363L..71D}  study is therefore unclear, unless the final donor star masses at the end of mass transfer in the latter authors' simulations are substantially higher than those in our simulations.}. However, higher-mass progenitors for the second-born NS are possible if spin-up of the first-born NS can occur during the common envelope phase of its companion. This is illustrated in Fig.~\ref{f4} where we show a scatterplot of the present-day orbital eccentricities and progenitor masses of the second-born NS in DNS binaries for the low kick model K20. In addition to the DNSs forming through the standard evolutionary channel, the figure also shows those that form without any mass transfer following the common envelope phase. Since we do not consider hypercritical accretion, the first-born NS in the latter group of systems is not spun-up in our models. 

Figure~\ref{f4} shows a clear dichotomy between low-$e$ systems associated with low-mass progenitors for the second-born NS, and high-$e$ systems associated with higher-mass progenitors for the second-born NS. The systems with low-mass progenitors all undergo stable mass transfer when the helium star evolves from core to shell helium burning and swells up to giant dimensions. At the start of the mass transfer phase these helium stars typically have masses below $3.5\,M_\odot$. Stars more massive than $3.5\,M_\odot$ do not expand much after exhausting their central helium supply, thereby avoiding any further mass transfer after the dynamically unstable common envelope phase. Since the eccentricities in the low kick model are predominantly determined by the mass lost during the SN explosion forming the second NS (and the subsequent evolution towards lower eccentricities due to gravitational wave emission), the separation of the second NS's progenitors in a low- and high-mass group causes in a dichotomy in the DNS orbital eccentricity range. While the observed sample of DNSs is still small, it is interesting to note that the observed DNS eccentricies show a lack of systems with $0.3 < e < 0.6$.  

Our limited understanding of NS accretion during common envelope evolution prohibits us from extending the simple spin-up model adopted in the case of dynamically stable mass transfer to dynamically unstable mass transfer. While it has been suggested that NSs in a common envelope may accrete enough matter to collapse into a black hole \citep{1986ApJ...308..755B, 1989ApJ...346..847C, 1993ApJ...411L..33C, 1995ApJ...440..270B, 1998ApJ...506..780B}, counter arguments claim rotation and jets may prevent significant accretion of matter \citep{1996ApJ...459..322C, 2000ApJ...532..540A}. Until significant improvements are made in our understanding of common envelope evolution, we are therefore left to speculate on the spin-up of the first-born NSs in the high-$e$ systems in Fig.~\ref{f4}. In this context, we note that if NSs are able to accrete during common envelope phases and if the amount of matter accreted decreases with increasing mass of the donor star (cf. Fig~\ref{f2}), the group of high-$e$ systems shown in Fig.~\ref{f4} can be expected to form an extension of the $P_{\rm spin}$--$e$ correlation found for low-$e$ systems in the right-hand panel of Fig.~\ref{f3}. The presence of high-$e$ systems in the observed $P_{\rm spin}$--$e$ relation can be interpreted as support for a common envelope model satisfying these assumptions.

\section{Discussion and conclusions}

We investigated the role of spin-kick alignment in establishing a correlation between the spin period of the first-born NS and the orbital eccentricity of DNSs using the StarTrack binary population synthesis code and a simple prescription for the spin-up of a NS due to mass accretion. For DNSs forming through the standard evolutionary channel, spin-kick alignment is compatible with the observed $P_{\rm spin}$--$e$ relation, but does not alleviate the requirement for low kick velocities proposed in previous population synthesis studies based on isotropic kick distributions. This is puzzling considering the stringent lower limits on the kick velocities imparted to the second-born NS in PSR\,B1534+12 and PSR\,B1913+16.  

Moreover, if low kicks are imparted to the second-born NS in DNSs, the standard formation channel cannot explain the formation of the high-eccentricity systems PSR\,B1913+16 and PSR\,J1811-1736. A possible resolution would be a dichotomous formation channel where low-eccentricty DNSs are formed through the standard formation channel with low kicks imparted to the second-born NS, and high-eccentricity systems are formed through a formation channel where no mass transfer occurs after the common envelope phase of the second NS's progenitor and  "normal" kicks typical of isolated radio pulsars are imparted to the second-born NS. The low-eccentricity systems on the $P_{\rm spin}$--$e$ relation can be formed with either polar or isotropic kicks. However, to obtain a $P_{\rm spin}$--$e$ relation at high eccentricities, some spin-kick alignment is required to counteract the increased spread in the post-SN orbital eccentricities introduced by the larger kicks. We will explore this in the continuation of this investigation.


\begin{theacknowledgments}
This work is partially supported by a Packard Foundation Fellowship, a NASA BEFS grant (NNG06GH87G), and a NSF CAREER grant (AST-0449558) to VK. 
\end{theacknowledgments}


\bibliographystyle{aipproc}   

\end{document}